\documentclass[aps,pra,twocolumn,groupedaddress,showpacs,superscriptaddress,scrartcl,amsart]{revtex4-1}
\usepackage{amssymb,amsmath,graphicx,epstopdf,hyperref,amsthm,dsfont,enumerate,bm,color}
\newtheorem{lemma}{Lemma}

\begin{document}
\interfootnotelinepenalty=10000

\title{Practical and unconditionally secure spacetime-constrained oblivious transfer}

\author{Dami\'an Pital\'ua-Garc\'ia}
\email{D.Pitalua-Garcia@damtp.cam.ac.uk}
\affiliation{IRIF, Unversit\'e Paris Diderot, Paris, France}
\thanks
{Present address: 
Centre for Quantum Information and Foundations, DAMTP, Centre for Mathematical Sciences, University of Cambridge, Wilberforce Road, Cambridge, CB3 0WA, U.K.}

\author{Iordanis Kerenidis}
\affiliation{IRIF, CNRS, Unversit\'e Paris Diderot, Paris, France}


\begin{abstract}
Spacetime-constrained oblivious transfer (SCOT) extends the fundamental primitive of oblivious transfer to Minkowski space. SCOT and location oblivious data transfer (LODT) are the only known cryptographic tasks with classical inputs and outputs for which unconditional security needs both quantum theory and relativity. We give an unconditionally secure SCOT protocol that, contrasting previous SCOT and LODT protocols, is practical to implement with current technology, where distant agents need only communicate classical information, while quantum communication occurs at a single location. We also show that our SCOT protocol can be used to implement unconditionally secure quantum relativistic bit commitment.
\end{abstract}


\maketitle
\section{Introduction}
\label{section1}

The two main theories in physics are quantum theory and relativity. Relativistic quantum cryptography \cite{K99,K99.2,K12.1} investigates the security of cryptographic protocols based on the physical principles of these theories. On the one hand, the causality of Minkowski spacetime -- and of the approximately-Minkowski spacetime near the Earth surface -- implies that physical systems and information cannot travel faster than light. On the other hand, quantum information cannot be cloned \cite{WZ82,D82}, non-orthogonal quantum states cannot be perfectly distinguished, quantum entanglement is monogamous \cite{T04}, and there are quantum correlations that violate Bell inequalities \cite{Bell}.

Some cryptographic primitives can be implemented with unconditional security based only on the principles of quantum theory, for example, quantum key distribution \cite{BB84,E91,BHK05}, private randomness expansion \cite{C09,PAMBMMOHLMM10,ColbeckKent11}, and weak coin tossing with arbitrarily small bias \cite{M07}. However, other important cryptographic tasks are impossible to implement with unconditional security based only on quantum theory, like bit commitment \cite{M97,LC97}, strong coin tossing with arbitrarily small bias \cite{LC98,Kitaev02,ABDR04}, oblivious transfer \cite{L97,R02,C07} and more general two-party classical computation \cite{L97,C07,BCS12} -- although there are quantum protocols with strictly better security than classical protocols for bit commitment \cite{CK11}, strong coin tossing \cite{CK09} and oblivious transfer \cite{CKS13}. Nevertheless, Kent showed that considering the causality of Minkowski spacetime allows us to obtain unconditionally secure protocols for strong coin tossing \cite{K99.2} and bit commitment \cite{K99,K12}, even if quantum theory is not considered.

It is interesting to ask whether there are cryptographic primitives whose unconditional security requires both quantum theory and relativity. While unconditionally secure protocols for oblivious transfer remain impossible to implement even with relativistic quantum protocols \cite{CK06,C07}, unconditionally secure protocols for its variants location-oblivious data transfer (LODT) \cite{K11.3} and spacetime-constrained oblivious transfer (SCOT) \cite{PG15.1} exist in relativistic quantum settings. However, unconditionally secure LODT and SCOT are impossible when based only on quantum theory, or in relativity. Only the combined properties of quantum information and Minkowski causality can allow for unconditionally secure LODT and SCOT protocols \cite{K11.3,PG15.1}. As far as we know, LODT and SCOT are the only known cryptographic tasks with classical inputs and outputs whose unconditional security necessitates both quantum theory and relativity.

LODT and SCOT are relativistic extensions of oblivious transfer. The primitive of one-out-of-two oblivious transfer \cite{EGL85} is informally defined as follows. Alice has two bits $x_0,x_1$ and Bob has a bit $b$. At the end of the protocol, Bob must learn $x_b$ but not $x_{\bar{b}}$, and Alice must not learn $b$. While somewhat abstract, oblivious transfer is related to private information retrieval and is a universal primitive, meaning that any secure computation between Alice and Bob can be reduced to this simple primitive \cite{K88}. In LODT \cite{K11.3}, Alice transfers a message to Bob at a random spacetime location that neither Alice nor Bob can determine in advance, and the location where Bob receives the message remains unknown to Alice. In SCOT \cite{PG15.1}, Alice has two input messages $x_0,x_1$,and  Bob has an input bit $b$; Bob obtains $x_b$ in $R_b$, but he cannot obtain $x_{\bar{b}}$ in $R_{\bar{b}}$, where $R_0$ and $R_1$ are fixed spacelike separated spacetime regions.

We motivate SCOT by the following scenario. Alice wants to give access to Bob to one of two computers: $\mathcal{C}_0$ or $\mathcal{C}_1$.
The computer $\mathcal{C}_i$ can only be accessed by Bob if he inputs the password $\bold{x}_i$ in the spacetime region $R_i$, for $i\in\{0,1\}$.
Each computer reveals a different database to Bob, or allows Bob to perform a different computational process. Bob does not want to reveal which computer he wants to access and Alice needs to make sure that Bob does not get access to both computers. SCOT guarantees that Bob can access one of the computers, without Alice knowing which one. Moreover, Alice is certain that, for a time interval, Bob cannot access the second computer too. Thus, by changing the access passwords before this time interval expires, the protocol guarantees that Bob can only access one of the computers. As mentioned above, oblivious transfer is a very strong cryptographic primitive \cite{K88}. SCOT is a strong cryptographic primitive too, as we show below that it implements unconditionally secure relativistic bit commitment. 

Although there are unconditionally secure protocols for LODT \cite{K11.3} and SCOT \cite{PG15.1}, these are not practical to implement with current technology. On the one hand, the LODT protocol of Ref. \cite{K11.3} requires preparation of two qudits in a maximally entangled state and their secure transmission over long distances. On the other hand, the SCOT protocol of Ref. \cite{PG15.1} requires preparation of Bennett-Brassard 1984 (BB84) \cite{BB84} qubits and their secure transmission over long distances -- as long as hundreds of kilometers if the time length of the regions $R_i$ is of only a few milliseconds -- and, if Bob does not have quantum memories to store the received qubits, the preparation and transmission of a large number of BB84 states within a very short time -- not greater than the time length of the regions $R_i$.

In this paper we introduce an unconditionally secure SCOT protocol that is practical to implement with current technology: it requires preparation, transmission and measurement of BB84 qubits at a single location, and secure transmission of classical information over long distances, but it does not require quantum memories or long-distance quantum communication. Furthermore, the quantum communication in our protocol can take an arbitrarily long time, while still giving Bob the freedom to chose his input $b$ only slightly in the past of $R_0$ and $R_1$. This is the first practical and unconditionally secure protocol of a cryptographic task whose unconditional security necessitates both quantum theory and relativity. Our SCOT protocol can also be used to implement unconditionally secure relativistic quantum bit commitment.

This paper is organized as follows. In Sec. \ref{section2} we describe the setting of relativistic quantum cryptography that applies to our protocol, we define the task of SCOT, and we provide some mathematical notation. We present our SCOT protocol in Sec. \ref{section3}. We show in Sec. \ref{section4} that our protocol is unconditionally secure. Some extensions of our protocol are discussed in Sec. \ref{section5}, in particular to deal with losses and errors. We discuss in Sec. \ref{section6} how our SCOT protocol can be used to implement unconditionally secure quantum relativistic bit commitment. We conclude in Sec. \ref{section7}.


\section{Preliminaries}
\label{section2}
\subsection{The setting}

The general model of relativistic quantum cryptography was introduced by Kent \cite{K99,K99.2,K12.1}. Here, we detail the specific setting of our protocol. Alice and Bob live in Minkowski -- or approximately Minkowski -- spacetime. Alice (Bob) consists of three different agents, denoted by $\mathcal{A}$, $\mathcal{A}_0$, $\mathcal{A}_1$ ($\mathcal{B}$, $\mathcal{B}_0$, $\mathcal{B}_1$), each controlling a secure laboratory adjacent to the space location $L$, $L_0$, $L_1$, having finite spatial extension with set of three-dimensional coordinates $\Delta L^{\text{A}},\Delta L_0^{\text{A}},\Delta L_1^{\text{A}}$ ($\Delta L^{\text{B}},\Delta L_0^{\text{B}},\Delta L_1^{\text{B}}$), respectively. Alice and Bob agree on a particular reference frame $\mathcal{F}$ with spacetime coordinates $(t,x,y,z),$ where the first entry is temporal, and where we set units in which the speed of light through vacuum is unity. Here we consider $L=(0,0,0)$ and $L_i=(-(-1)^i h,0,0)$, with the maximum length extension of any laboratory being much smaller than $h$, for $i\in\{0,1\}$.

We assume the processing of classical and quantum information within each laboratory to be secure and without errors, and the transmission of classical or quantum systems between laboratories being free of errors and losses. The communication between Alice's (Bob's) agents is done through authenticated and secure classical channels, which can be implemented with previously shared secret keys. There are classical channels between Alice's and Bob's adjacent agents, i.e. between the pairs $(\mathcal{A},\mathcal{B}), (\mathcal{A}_0,\mathcal{B}_0)$ and $(\mathcal{A}_1,\mathcal{B}_1)$. There is a single quantum channel between $\mathcal{A}$ and $\mathcal{B}$. Note that the only quantum communication is between agents at the same location,  an important difference with the setting in Ref. \cite{PG15.1}.


\subsection{Definition of SCOT}

As mentioned above, SCOT \cite{PG15.1} is a relativistic variant of one-out-of-two oblivious transfer \cite{EGL85}, defined as follows. Alice and Bob agree on two spacelike separated spacetime regions $R_0$ and $R_1$. Bob's agent $\mathcal{B}$ inputs $b \in \{0,1\}$ in the intersection of the causal pasts of all the spacetime points of $R_0$ and $R_1$.
For $i\in\{0,1\}$, Alice's agent $\mathcal{A}_i$ inputs $\bold{x}_i\in\{0,1\}^n$ in the causal past of a spacetime point $Q_i$ of $R_i$ -- previously agreed by Alice and Bob. The goal of the SCOT protocol is that $\mathcal{B}_b$ obtains $\bold{x}_b$ in $R_b$. A SCOT protocol is secure against Alice if she cannot learn $b$ anywhere in spacetime; it is secure against Bob if the probability that $\mathcal{B}_0$ obtains $\bold{x}_0$ in $R_0$ and $\mathcal{B}_1$ obtains $\bold{x}_1$ in $R_1$ goes to zero as a function of some security parameter, which here we take to be the parameter $n$ \cite{PG15.1}.

Here we consider $R_i=\{(t,\bold{l})\vert h\leq t\leq h+\Delta h, \bold{l}\in \Delta L_i^{\text{B}}\}$ and $Q_i=(h,-(-1)^ih,0,0)$, for $i\in\{0,1\}$, where $h>0$ and $\Delta h>0$. We note that $R_i$ includes the spacetime point $Q_i$, for $i\in\{0,1\}$. By definition of SCOT, $R_0$ and $R_1$ must be spacelike separated; hence, $\Delta h$ must be smaller than the shortest time that light takes to travel between $\Delta L_0^{\text{B}}$ and $\Delta L_1^{\text{B}}$, requiring $\Delta h<2h$. $\mathcal{B}$ generates $b$ within his laboratory in the past light cone of the spacetime point $P=(0,0,0,0)$. For $i\in\{0,1\}$, $\mathcal{A}_i$ inputs $\bold{x}_i$ within her laboratory in the past light cone of $Q_i$.

\subsection{Notation}

We define $[n]=\{1,2,\ldots,n\}$. We denote the $j$th bit entry of a string $\bold{a}\in\{0,1\}^n$ by $a^j$, for $j\in [n]$. When applied to bits (bit strings) $\oplus$ denotes (bitwise) sum modulo 2. We denote the complement of a bit $a$ by $\bar{a}=a\oplus 1$, and of a bit $a^j$ by $\bar{a}^j$.
$\mathcal{D}_0=\bigl\{\lvert 0\rangle,\lvert 1\rangle\bigr\}$ and $\mathcal{D}_1=\bigl\{\lvert \hat{0}\rangle,\lvert \hat{1}\rangle\bigr\}$ denote the computational and Hadamard bases, respectively, where $\lvert \hat{a}\rangle=\frac{1}{\sqrt{2}}\bigl(\lvert 0\rangle+(-1)^{a}\lvert 1\rangle\bigr)$, for $a\in\{0,1\}$.

\section{A practical and unconditionally secure SCOT protocol}
\label{section3}

We present an unconditionally secure SCOT protocol, consisting of two main stages. Stage I includes quantum communication between the laboratories adjacent to location $L$ and can be completed within any finite time in the past light cone of $P$. Stage II includes fast classical processing and communication between Alice's and Bob's agents at the locations $L$, $L_0$ and $L_1$, as well as classical communication between Alice's (Bob's) agents at the locations $L$ and $L_i$, for $i\in\{0,1\}$. Steps 1--6 take place in the past of $P$. Our protocol is illustrated in Fig. \ref{fig1}. 

\vspace{-0.3cm}
\subsection{Stage I}
\vspace{-0.3cm}

\begin{enumerate}
\item $\mathcal{A}$ generates  the strings  $\bold{r}_0,\bold{r}_1,\bold{s}\in\{0,1\}^n$ randomly and securely and sends copies to $\mathcal{A}_i$, who receives them in the past light cone of $Q_i$, for $i \in \{0,1\}$.

\item $\mathcal{A}$ prepares a set of $2n$ qubits $\bigl\{A_i^j\bigr\}_{i\in\{0,1\}}^{ j\in [n]}$ in BB84 states \cite{BB84}. For $j \in [n]$, the qubit pair $A^j=A_0^jA_1^j$ is prepared in the state
\begin{equation}
\label{1}
\bigl\lvert \psi_{r_0^jr_1^j}^{s^j}\bigr\rangle_{A_0^jA_1^j}=\bigl\lvert {r_0^j}\bigr\rangle_{A_{s^j}^j} \bigotimes \; \bigl \lvert \hat{r}_1^j\bigr\rangle_{A_{\bar{s}^j}^j}.
\end{equation}
We note that $s^j\in\{0,1\}$ indicates which qubit in the pair $A^j$ is prepared in the basis $\mathcal{D}_0$ and which one in $\mathcal{D}_1$. The qubits prepared in the basis $\mathcal{D}_i$ encode the string $\bold{r}_i$, for $i \in\{0,1\}$. For $i \in \{0,1\}$ and $j \in [n]$, $\mathcal{A}$ sends the qubits $A^j_i$ with their labels $i,j$ to $\mathcal{B}$.

\item Before receiving the qubits from $\mathcal{A}$, $\mathcal{B}$ randomly chooses a bit $c$. For $i \in \{0,1\}$ and $j \in [n]$, $\mathcal{B}$ measures the qubits $A^j_i$ in the basis $\mathcal{D}_{c}$, obtaining the bit outcomes $d^j_i$, defining $\bold{d}_i=(d^1_i,d^2_i,\ldots,d^{n}_i)$. For $i\in\{0,1\}$, $\mathcal{B}$ sends $c$, $\bold{d}_0$ and $\bold{d}_1$ to $\mathcal{B}_i$, with reception in the past of $Q_i$.

\end{enumerate}

\vspace{-0.3cm}
\subsection{Stage II}
\vspace{-0.3cm}

\begin{enumerate}
\setcounter{enumi}{3}
\item $\mathcal{B}$ generates his bit input $b$ and sends the bit $b'=c\oplus b$ to $\mathcal{A}$.

\item For $i\in\{0,1\}$, $\mathcal{B}$ sends $b$ to $\mathcal{B}_i$, who receives it in the past of $Q_i$.

\item For $i\in\{0,1\}$, $\mathcal{A}$ sends $b'$ to $\mathcal{A}_i$, who receives it in the past of $Q_i$.

\item For $i \in \{0,1\}$, $\mathcal{A}_i$ generates $\bold{x}_i$ within her laboratory in the past of $Q_i$, and gives $\bold{t}_i=\bold{r}_{i\oplus b'}\oplus \bold{x}_i$ and $\bold{s}$ to $\mathcal{B}_i$ at $Q_i$.

\item Within the region $R_b$, $\mathcal{B}_b$ uses $\bold{s}$, $\bold{d}_0$, $\bold{d}_1$, and $c$ to obtain $(d_{s^1\oplus c}^1,\ldots,d_{s^{n}\oplus c}^{n})$, which equals $\bold{r}_c$. Then, $\mathcal{B}_b$ outputs $\bold{y}_{b}=\bold{r}_c\oplus\bold{t}_b$, which equals $\bold{x}_b$.
\end{enumerate}

\begin{figure}
\includegraphics{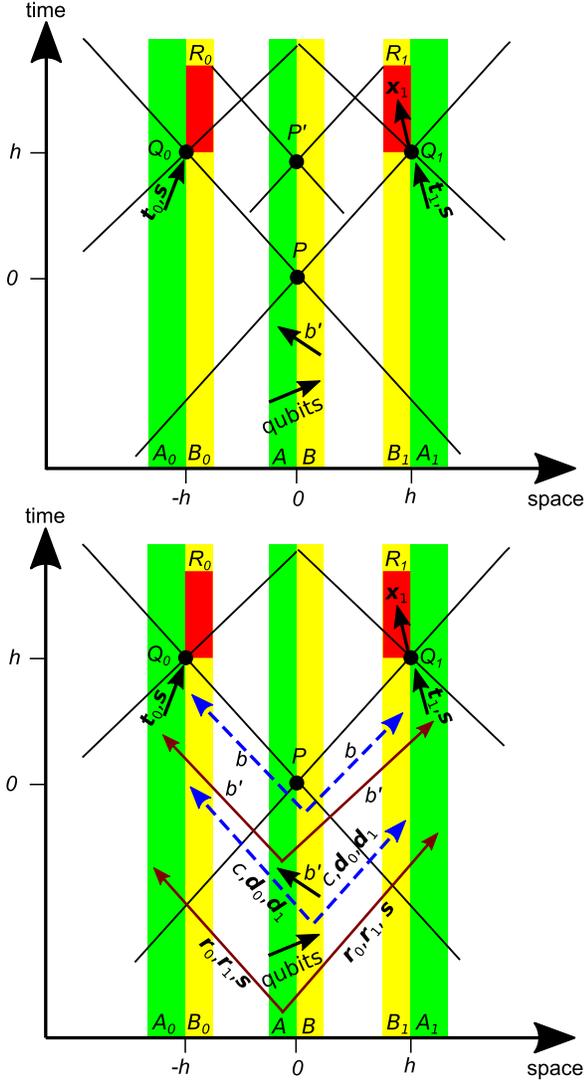}
 \caption{\label{fig1} Illustration of the introduced SCOT protocol in a two-dimensional spacetime diagram in the frame $\mathcal{F}$ of Minkowski spacetime. The world lines of the laboratories of Alice's agents $\mathcal{A}$, $\mathcal{A}_0$, $\mathcal{A}_1$ (green rectangles) and of the laboratories of Bob's agents $\mathcal{B}$, $\mathcal{B}_0$, $\mathcal{B}_1$ (yellow rectangles) are indicated. The small dots represent the spacetime points $P=(0,0,0,0)$, $Q_0=(h,-h,0,0)$, $Q_1=(h,h,0,0)$ and $P'$. The thin solid diagonal lines represent light rays, defining the light cones of $P$, $Q_0$, $Q_1$ and $P'$. The spacetime regions $R_i$, where Bob's agents must obtain Alice's inputs $\bold{x}_i$ correspond to the small red rectangles, for $i\in\{0,1\}$. The case $b=1$ is illustrated. Top: Only communication from Alice's (Bob's) agents to Bob's (Alice's) agents and Bob's output are illustrated (short black arrows with thick solid lines). $P'$ is the spacetime point with the greatest time coordinate in $\mathcal{F}$ that lies within the intersection of the past light cones of a point in $R_0$ and a point in $R_1$. When the SCOT protocol is used to implement bit commitment, Alice is guaranteed that Bob was committed to $b$ from $t'$, which is the time coordinate of $P'$. Bottom: All communication processes, including those among Alice's agents (long maroon arrows with thin solid lines) and among Bob's agents (long blue arrows with dashed lines), are illustrated.}
\end{figure}

\subsection{Comments and variations}

We note that $\mathcal{B}$ has the freedom to choose $b$ after all the received qubits have been measured. This is useful in situations where the quantum communication takes a long time (e.g., several hours), while Bob wants to decide which message to obtain within a short time (e.g., a fraction of a second) in the past of $R_0$ and $R_1$. Furthermore, Alice has the freedom to generate her inputs in real time; i.e., $\mathcal{A}_i$ can generate $\bold{x}_i$ slightly in the past of $Q_i$, for $i\in\{0,1\}$.

Different variations of the above protocol can be considered. For example, in one variation $\mathcal{B}$ does not send $b$ to his agents and $\mathcal{B}_i$ acts assuming that $b=i$, for $i\in\{0,1\}$. In another variation, $\mathcal{A}$ generates $\bold{x}_0$ and $\bold{x}_1$ in the past of $P$ and sends $\bold{x}_i$ to $\mathcal{A}_i$, who receives it in the past of $Q_i$, for $i\in\{0,1\}$.

\section{Security analysis}
\label{section4}
\subsection{Security against Alice}

The only communication from Bob to Alice consists in the bit $b'=c\oplus b$. Since $c$ is random and Bob's laboratories and classical channels are secure, Alice cannot obtain any information about $b$. Thus, our protocol is perfectly secure against Alice.

\subsection{Security against Bob}

Security against Bob follows from Minkowski causality and the monogamy of entanglement. In a general cheating strategy by Bob, agent $\mathcal{B}$ introduces an ancilla $E$ of arbitrary finite dimension in some fixed pure state $\lvert \chi \rangle$, and applies a unitary operation $U$ on $AE$ to obtain
\begin{equation}
\label{2}
\lvert \Phi_{\bold{r}_0\bold{r}_1}^s\rangle_{B_0B_1B'} =U_{AE} \lvert \Psi_{\bold{r}_0\bold{r}_1}^s\rangle_{A}\lvert \chi\rangle_E ,
\end{equation}
where $A=A^1_0A^1_1\cdots A^{n}_0A^{n}_1$, $AE=B_0B_1B'$, the systems $B_0,B_1,B'$ have arbitrary finite dimensions, and where
\begin{equation}
\label{3}
\bigl\lvert \Psi_{\bold{r}_0\bold{r}_1}^s\bigr\rangle_{A}=\bigotimes_{j=1}^{n} \bigl\lvert \psi_{r_0^jr_1^j}^{s^j}\bigr\rangle_{A_0^jA_1^j}
\end{equation}
is the state that $\mathcal{B}$ receives from $\mathcal{A}$.
Then, $\mathcal{B}$ applies a projective measurement $\{R^{b'}\}_{b'\in\{0,1\}}$ on $B'$ and sends the outcome $b'$ to $\mathcal{A}$. For $i \in \{0,1\}$, $\mathcal{B}$ sends $b'$ and $B_i$ to $\mathcal{B}_i$. $\mathcal{B}_i$, after receiving $\bold{s}$ from $\mathcal{A}_i$ at $Q_i$, applies a projective measurement $M_{i\bold{s}b'}=\{\Pi_{i \bold{s}b'}^{\bold{e}_i}\}_{\bold{e}_i\in\{0,1\}^n}$ on $B_i$ and obtains the outcome $\bold{e}_i$. Given that $\bold{t}_i=\bold{r}_{i\oplus b'}\oplus\bold{x}_i$, the goal of the measurement $M_{i\bold{s}b'}$ is to obtain a string $\bold{e}_i$ that equals $\bold{r}_{i\oplus b'}$ with high probability, so that $\mathcal{B}_i$ outputs $\bold{e}_i\oplus\bold{t}_i$, which equals $\bold{x}_i$ with high probability. Thus, Bob's cheating probability is $p_n=P(\bold{e}_0=\bold{r}_{b'},\bold{e}_1=\bold{r}_{\bar{b}'})$, which is given by
\begin{equation}
\label{4}
p_n=\frac{1}{2^{3n}}\sum_{\bold{r}_0,\bold{r}_1,\bold{s},b'}\langle \Phi_{\bold{r}_0\bold{r}_1}^\bold{s}\rvert \Pi_{0\bold{s}b'}^{\bold{r}_{b'}}\otimes \Pi_{1\bold{s}b'}^{\bold{r}_{\bar{b}'}}\otimes R^{b'}\rvert \Phi_{\bold{r}_0\bold{r}_1}^\bold{s}\rangle.
\end{equation}

Since the regions $R_0$ and $R_1$ at which Bob's agents $\mathcal{B}_0$ and $\mathcal{B}_1$ must produce their outputs are spacelike separated, it follows from Minkowski causality that, as given by (\ref{4}), $\mathcal{B}_0$ and $\mathcal{B}_1$ must obtain their outputs from measurements on disjoint systems $B_0$ and $B_1$, which could have only interacted in the intersection of the past light cones of at least one point in $R_0$ and one point in $R_1$, i.e in the past light cone of $P'$ (see Fig. \ref{fig1}). Then, from the monogamy of entanglement, the measurement outcomes obtained by $\mathcal{B}_0$ and $\mathcal{B}_1$ cannot both be very correlated to Alice's inputs $\bold{r}_0$ and $\bold{r}_1$, which implies an upper bound on $p_n$. We show below that for any cheating strategy by Bob we have
\begin{equation}
\label{5}
p_n\leq \Bigl( \frac{1}{2}+\frac{1}{2\sqrt{2}}\Bigr)^n.
\end{equation}
Thus, $p_n\rightarrow 0$ exponentially with $n$, meaning that our protocol is unconditionally secure.

It remains as an open question whether the security bound (\ref{5}) is tight. We describe in section \ref{cheat} the best cheating strategy by Bob that we have found, which achieves $p_n=\bigl(\frac{3}{4}\bigr)^n$.

\subsubsection{Proof of bound (\ref{5})}

The main technical part of this paper is the proof of bound (\ref{5}). Our proof is a variation of the proof of a bound for the monogamy-of-entanglement game given in Ref. \cite{TFKW13}.
First, we note that it is equivalent in our protocol if agent $\mathcal{A}$ prepares in the past of $P$ the state
\begin{equation}
\label{9}
\lvert \Phi\rangle_{CA}=\bigotimes_{j=1}^{n}\Bigl(\lvert \Phi^+\rangle_{C_0^jA_0^j}\otimes\lvert\Phi^+\rangle_{C_1^jA_1^j}\Bigr),
\end{equation}
where $C=C^1C^2\cdots C^{n}$, $C^j=C_0^jC_1^j$ and $\lvert \Phi^+\rangle=\frac{1}{\sqrt{2}}\bigl(\lvert 0\rangle\lvert 0\rangle+\lvert 1\rangle\lvert 1\rangle\bigr)$. Then, in the past of $P$, $\mathcal{A}$ sends $A$ to $\mathcal{B}$ and measures $C^j$ in the basis $\bigl\{\bigl\lvert \psi_{r_0r_1}^{s^j}\bigr\rangle\bigr\}_{r_0,r_1\in\{0,1\}}$ for $j\in[n]$. With probability $\frac{1}{4}$, $\mathcal{A}$ measures $C^j$ in the state $\bigl\lvert \psi_{r_0^jr_1^j}^{s^j}\bigr\rangle_{C_0^jC_1^j}$ and the pair of qubits $A^j$ received by $\mathcal{B}$ collapse into the state $\bigl\lvert \psi_{r_0^jr_1^j}^{s^j}\bigr\rangle_{A_0^jA_1^j}$. Bob's unitary operation $U$ in his cheating strategy commutes with Alice's measurements. Thus, we can consider that the global system $CB_0B_1B'$, before Alice's and Bob's measurement are implemented, is in the state
\begin{equation}
\label{10}
\lvert \Psi\rangle_{CB_0B_1B'}=\bigl(\mathds{1}_C\otimes U_{B_0B_1B'}\bigr)\lvert \Phi\rangle_{CA}\lvert \chi\rangle_E,
\end{equation}
where we recall that $B_0B_1B'=AE$, $E$ is an ancilla, and $B_0$, $B_1$ and $B'$ have arbitrary finite Hilbert space dimensions. Then, $\mathcal{A}$ measures $C$ in the orthonormal basis $\bigl\{\bigl\lvert \Psi_{\bold{r}_0\bold{r}_1}^{\bold{s}}\bigr\rangle\bigr\}_{\bold{r}_0,\bold{r}_1\in\{0,1\}^n}$ according to her random value of $\bold{s}\in\{0,1\}^n$. With probability $\frac{1}{2^{2n}}$, $\mathcal{A}$ measures $\bigl\lvert \Psi_{\bold{r}_0\bold{r}_1}^{\bold{s}}\bigr\rangle_C$ and $B_0B_1B'$ projects into the state $\lvert \Phi_{\bold{r}_0\bold{r}_1}^s\rangle_{B_0B_1B'}$. Then, $\mathcal{B}$ applies the projective measurement $\{R^{b'}\}_{b'\in\{0,1\}}$ on $B'$ and obtains the bit outcome $b'$, which he sends to $\mathcal{A}$. $\mathcal{B}$ sends $b'$ and $B_i$ to $\mathcal{B}_i$, for $i\in\{0,1\}$. After receiving $\bold{s}$, agent $\mathcal{B}_i$ applies the projective measurement $M_{i\bold{s}b'}$ on $B_i$ at $Q_i$. Thus, Bob's cheating probability $p_n$ given by (\ref{4})
 equals
\begin{equation}
\label{11}
p_n=\frac{1}{2^{n}}\sum_{\bold{s}}\text{Tr} \bigl(T_{\bold{s}}\Psi\bigr),
\end{equation}
where $\Psi=\bigl(\lvert \Psi\rangle\langle \Psi\rvert\bigr)_{CB_0B_1B'}$ and 
\begin{equation}
\label{aa1}
T_{\bold{s}}=\sum_{b'=0}^1 \bigl(D_{\bold{s}}^{b'}\bigr)_{CB_0B_1}\otimes \bigl(R^{b'}\bigr)_{B'},
\end{equation}
with 
\begin{equation}
\label{12}
D_{\bold{s}}^{b'} = \sum_{\bold{r}_0,\bold{r}_1} \bigl( \bigl\lvert \Psi_{\bold{r}_0\bold{r}_1}^{\bold{s}}\bigr\rangle\bigl\langle\Psi_{\bold{r}_0\bold{r}_1}^{\bold{s}}\bigr\rvert\bigr)_{C}\otimes \bigl(\Pi_{0\bold{s}b'}^{\bold{r}_{b'}}\bigr)_{B_0}\otimes \bigl(\Pi_{1\bold{s}b'}^{\bold{r}_{\bar{b}'}}\bigr)_{B_1},
\end{equation}
for $b'\in\{0,1\}$.

We derive bound (\ref{5}) with the help of two lemmas of Ref. \cite{TFKW13}. Before stating these lemmas we provide some useful notation. We denote by $\mathcal{H}$ the Hilbert space of the global system $CB_0B_1B'$, which as said before is arbitrary but finite dimensional. We denote by $\mathcal{L}(\mathcal{H})$ and by $\mathcal{P}(\mathcal{H})$ the sets of linear operators and of positive semi-definite operators on $\mathcal{H}$, respectively. For $A,B\in\mathcal{L}(\mathcal{H})$, the expression $A\geq B$ means that $A-B\in\mathcal{P}(\mathcal{H})$. For $A\in\mathcal{L}(\mathcal{H})$, $\lVert A\rVert$ denotes the Schatten $\infty-$norm of $A$, which gives the largest singular value of $A$, and which coincides with its largest eigenvalue if $A\in\mathcal{P}(\mathcal{H})$.

\begin{lemma}{(Ref. \cite{TFKW13})}
\label{lemma1}
Let $A,B,L\in\mathcal{L}(\mathcal{H})$ such that $A^{\dagger}A\geq B^{\dagger}B$. Then, it holds that $\lVert AL \rVert \geq \lVert BL \rVert$.
\end{lemma}


\begin{lemma}{(Ref. \cite{TFKW13})}
\label{lemma2}
Let $D_1,D_2,\ldots,D_N\in \mathcal{P}(\mathcal{H})$, and let $\{s_k\}_{k\in [N]}$ be a set of $N$ mutually orthogonal permutations of $[N]$. Then
\begin{equation}
\biggl\lVert \sum_{i\in [N]} D_i\biggr \rVert\leq \sum_{k\in [N]}\max_{i\in [N]}\Bigl \lVert \sqrt{D_i}\sqrt{D_{s_k(i)}}\Bigr \rVert.
\end{equation}
\end{lemma}


It follows from Lemma \ref{lemma1} that for $A,A',B,B'\in\mathcal{P}(\mathcal{H})$ satisfying $A'\geq A$ and $B' \geq B$, it holds that $\lVert \sqrt{A'}\sqrt{B'}\rVert \geq \lVert \sqrt{A'}\sqrt{B}\lVert \geq \lVert \sqrt{A}\sqrt{B}\rVert$ \cite{TFKW13}. Thus, if $A,A',B,B'$ are projectors on $\mathcal{H}$ satisfying $A'\geq A$ and $B' \geq B$ then $\lVert A'B'\rVert \geq \lVert AB\rVert$. We use this property below.

To use Lemma \ref{lemma2}, we consider the set of permutations of $\bold{s}$ given by $\bold{s}\rightarrow \bold{s}_{\bold{k}}=\bold{s}\oplus\bold{k}$, for $\bold{k}\in\{0,1\}^n$. This is a set of $2^n$ mutually orthogonal permutations, that is,  $\bold{s}_{\bold{k}}\neq \bold{s}_{\bold{k}'}$ if $\bold{k}\neq\bold{k}'$ for all $\bold{s}\in\{0,1\}^n$. 

We have
\begin{eqnarray}
\label{13}
p_n&=&\frac{1}{2^n}\text{Tr}\Biggl(\sum_{\bold{s}}T_{\bold{s}}  \Psi\Biggr)\nonumber\\
&\leq&\frac{1}{2^n} \Biggl\lVert \sum_{\bold{s}} T_{\bold{s}} \Biggr\rVert\nonumber\\
&=&\frac{1}{2^n} \Biggl\lVert \sum_{b'=0}^1\sum_{\bold{s}} D_{\bold{s}}^{b'}\otimes R^{b'} \Biggr\rVert\nonumber\\
&=&\frac{1}{2^n}\max_{b'} \Biggl\lVert \sum_{\bold{s}} D_{\bold{s}}^{b'} \Biggr\rVert\nonumber\\
&\leq&\frac{1}{2^n}\max_{b'}\sum_{\bold{k}}\max_{\bold{s}}\Bigl\lVert D^{b'}_{\bold{s}}D^{b'}_{\bold{s}_{\bold{k}}}\Bigr\rVert,
\end{eqnarray}
where in the first line we used the linearity of the trace, in the second line we used the definition of the Schatten $\infty-$norm, in the third line we used (\ref{aa1}), in the fourth line we used that $\{R^0,R^1\}$ is a projective measurement, and in the last line we used Lemma \ref{lemma2} and the fact that $D_{\bold{s}}^{b'}$ and $D_{\bold{s}_{\bold{k}}}^{b'}$ are projectors.

We define the projectors
\begin{eqnarray}
\label{18}
F^{b'}_{\bold{s}}&=&\sum_{\bold{r}_0,\bold{r}_1 }\bigl( \bigl\lvert \Psi_{\bold{r}_0\bold{r}_1}^{\bold{s}}\bigr\rangle\bigl\langle\Psi_{\bold{r}_0\bold{r}_1}^{\bold{s}}\bigr\rvert\bigr)_{C}\otimes \bigl(\Pi_{0\bold{s}b'}^{\bold{r}_{b'}}\bigr)_{B_0}\otimes \mathds{1}_{B_1},\nonumber\\
G^{b'}_{\bold{s}_{\bold{k}}}&=&\sum_{\bold{r}_0,\bold{r}_1}\bigl( \bigl\lvert \Psi_{\bold{r}_0\bold{r}_1}^{\bold{s}_{\bold{k}}}\bigr\rangle\bigl\langle\Psi_{\bold{r}_0\bold{r}_1}^{\bold{s}_{\bold{k}}}\bigr\rvert\bigr)_{C}\otimes  \mathds{1}_{B_0}\otimes\bigl(\Pi_{1\bold{s}_{\bold{k}}b'}^{\bold{r}_{\bar{b}'}}\bigr)_{B_1},\nonumber\\
\end{eqnarray}
for $b'\in\{0,1\}$ and $\bold{s},\bold{k}\in\{0,1\}^n$. We see that they satisfy $D^{b'}_{\bold{s}}\leq F^{b'}_{\bold{s}}$ and $ D^{b'}_{\bold{s}_{\bold{k}}}\leq G^{b'}_{\bold{s}_{\bold{k}}}$. Thus, we have from Lemma \ref{lemma1} that
\begin{equation}
\label{14}
\lVert D^{b'}_{\bold{s}}D^{b'}_{\bold{s}_{\bold{k}}}\rVert^2\leq \lVert F^{b'}_{\bold{s}}G^{b'}_{\bold{s}_{\bold{k}}}\rVert ^2=\lVert F^{b'}_{\bold{s}}G^{b'}_{\bold{s}_{\bold{k}}}F^{b'}_{\bold{s}}\rVert,
\end{equation}
where the equality follows from the property $\lVert A\rVert^2=\lVert AA^{\dagger}\rVert=\lVert A^{\dagger}A\rVert$ for any $A\in\mathcal{L}(\mathcal{H})$ \cite{TFKW13} and from the fact that $F^{b'}_{\bold{s}}$ and $G^{b'}_{\bold{s}_{\bold{k}}}$ are projectors. We show in Appendix \ref{appa} that
\begin{equation}
\label{15}
\lVert F^{b'}_{\bold{s}}G^{b'}_{\bold{s}_{\bold{k}}}F^{b'}_{\bold{s}}\rVert= \Bigl(\frac{1}{2}\Bigr)^{\omega_\bold{k}},
\end{equation}
where $\omega_\bold{k}$ is the Hamming weight of $\bold{k}$. Thus, since for any $b'\in\{0,1\}$ and for a fixed $\bold{k}$, the value of $\lVert F^{b'}_{\bold{s}}G^{b'}_{\bold{s}_{\bold{k}}}F^{b'}_{\bold{s}}\rVert$ is the same for any $\bold{s}\in\{0,1\}^n$, we have from (\ref{13}), (\ref{14}) and (\ref{15}) that
\begin{equation}
\label{16}
p_n\leq \frac{1}{2^n}\sum_{\bold{k}} \Bigl(\frac{1}{\sqrt{2}}\Bigr)^{\omega_\bold{k}}.
\end{equation}
There are exactly $\bigl(\begin{smallmatrix} n \\ \omega\end{smallmatrix} \bigr)$ values of $\bold{k}\in\{0,1\}^n$ with Hamming weight $\omega$. Thus, from (\ref{16}), we obtain
\begin{eqnarray}
\label{17}
p_n&\leq&\frac{1}{2^n}\sum_{\omega=0}^{n}\Bigl(\begin{matrix}
  n \\  \omega \end{matrix}\Bigr)\Bigl(\frac{1}{\sqrt{2}}\Bigr)^\omega\nonumber\\
&=&\Bigl(\frac{1}{2}+\frac{1}{2\sqrt{2}}\Bigr)^n,
\end{eqnarray}
which is the claimed bound (\ref{5}).


\subsection{A cheating strategy by Bob without the use of any quantum channels or quantum memories}
\label{cheat}
We present a cheating strategy by Bob that achieves $p_n=\bigl(\frac{3}{4}\bigr)^n$. Interestingly, Bob's agents do not need any quantum memories or quantum channels to perform this strategy.

Bob's agent $\mathcal{B}$ gives the bit $b'=0$ to Alice's agent $\mathcal{A}$. Thus, Bob's agent $\mathcal{B}_i$ must obtain the string $\bold{r}_i$ in $R_i$, for $i\in\{0,1\}$. In the past of $P$, agent $\mathcal{B}$ applies a projective measurement on the pair of qubits $A^j$ in the orthonormal basis defined by the following states:
\begin{eqnarray}
\label{6}
\lvert \xi_{00}\rangle&=&\frac{\sqrt{3}}{2}\lvert \psi_{00}^0\rangle-\frac{1}{2\sqrt{3}}\bigl(-\lvert \psi_{01}^0\rangle+\lvert \psi_{10}^0\rangle+\lvert \psi_{11}^0\rangle\bigr)\nonumber\\
\lvert \xi_{01}\rangle&=&\frac{\sqrt{3}}{2}\lvert \psi_{01}^0\rangle-\frac{1}{2\sqrt{3}}\bigl(\lvert \psi_{00}^0\rangle-\lvert \psi_{10}^0\rangle+\lvert \psi_{11}^0\rangle\bigr)\nonumber\\
\lvert \xi_{10}\rangle&=&\frac{\sqrt{3}}{2}\lvert \psi_{10}^0\rangle-\frac{1}{2\sqrt{3}}\bigl(-\lvert \psi_{00}^0\rangle+\lvert \psi_{01}^0\rangle+\lvert \psi_{11}^0\rangle\bigr)\nonumber\\
\lvert \xi_{11}\rangle&=&-\frac{\sqrt{3}}{2}\lvert \psi_{11}^0\rangle-\frac{1}{2\sqrt{3}}\bigl(\lvert \psi_{00}^0\rangle+\lvert \psi_{01}^0\rangle+\lvert \psi_{10}^0\rangle\bigr).\nonumber\\
\end{eqnarray}
We denote the outcomes by $\bigl\lvert \xi_{l^j}\bigr\rangle$, where $l^j=(l_0^j,l_1^j)\in\{0,1\}^2$, for $j\in [n]$. Agent $\mathcal{B}$ sends the outcomes $l^j$ with their labels $j$ to his agent $\mathcal{B}_i$, who receives them in the past of $Q_i$, for $j\in[n]$ and $i\in\{0,1\}$. After receiving $\bold{s}$ at $Q_i$ from $\mathcal{A}_i$, agent $\mathcal{B}_i$ guesses that Alice prepared the pair of qubits $A^j$ in the state $\lvert \psi_{l^j}^0\rangle$ if $s^j=0$ or in the state $\bigl\lvert \psi_{f(l^j)}^{1}\bigr\rangle$ if $s^j=1$, where $f(00)=01$, $f(01)=11$, $f(10)=10$ and $f(11)=00$. Thus, $\mathcal{B}_i$ outputs $\bold{e}_i=(e_i^1,e_i^2,\ldots,e_i^n)$, with $e_i^j=l_i^j$ if $s^j=0$, or with $e_i^j$ being the $i$th entry of $f(l^j)$ if $s^j=1$. Bob's cheating probability $p_n=P(\bold{e}_0=\bold{r}_0,\bold{e}_1=\bold{r}_1)$ following this strategy is $p_n=(p)^n$, where
\begin{eqnarray}
\label{7}
p&=&\frac{1}{8}\Bigl(\bigl\lvert \bigl\langle \psi_{00}^0\big\vert \xi_{00}\bigr\rangle\bigr\rvert^2+\bigl\lvert \bigl\langle \psi_{01}^0\big\vert \xi_{01}\bigr\rangle\bigr\rvert^2+\bigl\lvert \bigl\langle \psi_{10}^0\big\vert \xi_{10}\bigr\rangle\bigr\rvert^2\nonumber\\
&&\quad\quad+\bigl\lvert \bigl\langle \psi_{11}^0\big\vert \xi_{11}\bigr\rangle\bigr\rvert^2+
\bigl\lvert \bigl\langle \psi_{00}^1\big\vert \xi_{11}\bigr\rangle\bigr\rvert^2+\bigl\lvert \bigl\langle \psi_{01}^1\big\vert \xi_{00}\bigr\rangle\bigr\rvert^2\nonumber\\
&&\quad\quad+\bigl\lvert \bigl\langle \psi_{10}^1\big\vert \xi_{10}\bigr\rangle\bigr\rvert^2+\bigl\lvert \bigl\langle \psi_{11}^1\big\vert \xi_{01}\bigr\rangle\bigr\rvert^2\Bigr).
\end{eqnarray}
From the definition (\ref{1}) for the states $\lvert \psi_{r_0r_1}^s\rangle$, we can easily derive the relations
\begin{eqnarray}
\label{8}
\bigl\lvert \psi_{00}^1\bigr\rangle&=&\frac{1}{2}\Bigl(\bigl\lvert \psi_{00}^0\bigr\rangle+\bigl\lvert \psi_{01}^0\bigr\rangle+\bigl\lvert \psi_{10}^0\bigr\rangle+\bigl\lvert \psi_{11}^0\bigr\rangle\Bigr)\nonumber\\
\bigl\lvert \psi_{01}^1\bigr\rangle&=&\frac{1}{2}\bigl(\bigl\lvert \psi_{00}^0\bigr\rangle+\bigl\lvert \psi_{01}^0\bigr\rangle-\bigl\lvert \psi_{10}^0\bigr\rangle-\bigl\lvert \psi_{11}^0\bigr\rangle\Bigr)\nonumber\\
\bigl\lvert \psi_{10}^1\bigr\rangle&=&\frac{1}{2}\bigl(\bigl\lvert \psi_{00}^0\bigr\rangle-\bigl\lvert \psi_{01}^0\bigr\rangle+\bigl\lvert \psi_{10}^0\bigr\rangle-\bigl\lvert \psi_{11}^0\bigr\rangle\Bigr)\nonumber\\
\bigl\lvert \psi_{11}^1\bigr\rangle&=&\frac{1}{2}\Bigl(\bigl\lvert \psi_{00}^0\bigr\rangle-\bigl\lvert \psi_{01}^0\bigr\rangle-\bigl\lvert \psi_{10}^0\bigr\rangle+\bigl\lvert \psi_{11}^0\bigr\rangle\Bigr).
\end{eqnarray}
Using (\ref{6}) -- (\ref{8}) it is straightforward to compute $p=\frac{3}{4}$. Thus, this cheating strategy by Bob achieves a success probability $p_n=\bigl(\frac{3}{4}\bigr)^n$, as claimed.

\section{Extensions of our SCOT protocol}
\label{section5}

We discuss possible extensions of the SCOT protocol of Sec. \ref{section3}. First, we note that in the SCOT protocol of Sec. \ref{section3} Bob is guaranteed that the pair of qubits $A^j$ encodes the bits $x_0^j$ and $x_1^j$ via the bits $r_0^j$ and $r_1^j$. In a straightforward extension of our protocol, among the $2n$ qubits that Alice prepares from the BB84 set ($n$ in the computational basis and $n$ in the Hadamard basis), she chooses randomly the positions of the qubits encoding the bits $r_i^j$, for $i\in\{0,1\}$ and $j\in [n]$. Clearly, this modification in the protocol cannot make it easier for Bob to cheat; hence, unconditional security remains. We leave as an open question whether this extended protocol achieves strictly better security against Bob.

Secondly, the SCOT protocol of Sec. \ref{section3} considers the ideal situation in which there are not any errors during the protocol and that there are not any losses of the transmitted quantum systems. This protocol is straightforwardly extended to deal with these problems.

\subsection{Dealing with losses}

To deal with losses, $\mathcal{B}$ reports to $\mathcal{A}$ in the past of $P$ a set $\mathcal{S}\subseteq [n]$ of labels $j$ of two-qubit systems $A^j=A^j_0A^j_1$ for which both $A^j_0$ and $A^j_1$ activate a detection. The protocol continues as the original one, but with input strings $\bold{s}^{\mathcal{S}}$, $\bold{r}_0^{\mathcal{S}}$, $\bold{r}_1^{\mathcal{S}}$, $\bold{d}_0^{\mathcal{S}}$, $\bold{d}_1^{\mathcal{S}}$, $\bold{x}_0^{\mathcal{S}}$, $\bold{x}_1^{\mathcal{S}}$ instead of $\bold{s}$, $\bold{r}_0$, $\bold{r}_1$, $\bold{d}_0$, $\bold{d}_1$, $\bold{x}_0$, $\bold{x}_1$, where $\bold{a}^{\mathcal{S}}\in\{0,1\}^N$ denotes the restriction of $\bold{a}\in\{0,1\}^n$ in $\mathcal{S}$, and where $N=\lvert \mathcal{S}\rvert$. 
Security against Bob follows from bound (\ref{5}) but with security parameter $N$ instead of $n$. Since Bob's agent $\mathcal{B}$ gives Alice the bit $b'=c\oplus b$, in order to guarantee security against Alice, $\mathcal{B}$ must make sure that announcing the detection events does not leak any information about the measurement choice $c$, which implies that no information about Bob's input $b$ is given to Alice. For this, $\mathcal{B}$ must experimentally verify that the detection probabilities of the two-qubit systems $A^j$ received from $\mathcal{A}$ are independent of whether they are measured in the basis $\mathcal{D}_0\otimes\mathcal{D}_0$ or in the basis $\mathcal{D}_1\otimes \mathcal{D}_1$.

\subsection{Dealing with errors}
In order to deal with noise in the system we extend the setting of SCOT. As before, in the past light cone of $Q_i$, Alice's agent $\mathcal{A}_i$ has the input $\bold{x}_i\in\{0,1\}^n$, for $i\in\{0,1\}$. In the past light cone of $P$, Bob's agent $\mathcal{B}$ has as an input $b \in \{0,1\}$. However, now, the goal of the protocol is that Bob obtains a string $\bold{y}_b\in\{0,1\}^n$ in $R_b$ such that $\bold{y}_b$ differs from $\bold{x}_b$ in at most $\gamma n$ bit entries, for some $\gamma\in (0,1)$ agreed in advanced by Alice and Bob. As before, security against Alice means that she cannot learn $b$ anywhere in spacetime. Security against Bob means that Bob cannot obtain a string $\bold{y}_0$ in $R_0$ and a string $\bold{y}_1$ in $R_1$, such that $\bold{y}_i$ differs from $\bold{x}_i$ in at most $\gamma n$ bit entries for both $i=0$ and $i=1$. According to the motivation of SCOT outlined in the Introduction, this means that Bob cannot access both computers, one being accessed in $R_0$ and the other one being accessed in $R_1$, where the corresponding computer can be accessed if the password provided by Bob's agent differs from the original password given by Alice's agent in at most $\gamma n$ bit entries. It can be shown that our protocol remains unconditionally secure as long as $\gamma$ is small enough. In particular, we show below that our protocol remains secure if $\gamma< 0.015$. 

Bob's cheating probability in the extended setting, the probability that both agents $\mathcal{B}_0$ and $\mathcal{B}_1$ output valid passwords, is
\begin{equation}
\label{a1}
p_n^\gamma=\frac{1}{2^{n}}\sum_{\bold{q}\in\mathcal{Q}_n^\gamma}\sum_{\bold{s}}\text{Tr} \bigl(T_{\bold{s}}^{\bold{q}}\Psi\bigr),
\end{equation}
where $\Psi=\bigl(\lvert \Psi\rangle\langle \Psi\rvert\bigr)_{CB_0B_1B'}$, with $\lvert \Psi\rangle_{CB_0B_1B'}$ given by (\ref{10}), and $T_{\bold{s}}^{\bold{q}}$ is a projector given by
\begin{equation}
\label{aa2}
T_{\bold{s}}^{\bold{q}}=\sum_{b'=0}^1 \bigl(D_{\bold{s}}^{\bold{q}b'}\bigr)_{CB_0B_1}\otimes \bigl(R^{b'}\bigr)_{B'},
\end{equation}
with 
\begin{eqnarray}
\label{a2}
D_{\bold{s}}^{\bold{q}b'}&=&\sum_{\bold{r}_0,\bold{r}_1}\Bigl[\bigl(\bigl\lvert\Psi_{\bold{r}_0\bold{r}_1}^{\bold{s}}\bigr\rangle\bigl\langle\Psi_{\bold{r}_0\bold{r}_1}^{\bold{s}}\bigr\rvert\bigr)_{C}\otimes \bigl(\Pi_{0\bold{s}b'}^{\bold{r}_{b'}\oplus \bold{q}_0}\bigr)_{B_0}\times\nonumber\\
&&\quad\quad\times\otimes \bigl(\Pi_{1\bold{s}b'}^{\bold{r}_{\bar{b}'}\oplus\bold{q}_1}\bigr)_{B_1}\Bigr];
\end{eqnarray}
and where $\bold{q}=(\bold{q}_0,\bold{q}_1)\in \mathcal{Q}_n^\gamma\equiv \bigl\{(\bold{q}_0,\bold{q}_1)\big\vert \bold{q}_i\in\{0,1\}^n, \mathcal{W}(\bold{q}_i)\leq \gamma n \text{ for } i=0,1\bigr\}$, with $\mathcal{W}(\bold{q}_i)$ denoting the Hamming weight of $\bold{q}_i$, for fixed  $\gamma\in [0,1]$. We show in Appendix \ref{appb} that
\begin{equation}
\label{a3}
p_n^\gamma \leq \bigl \lvert \mathcal{Q}_n^\gamma \bigr\rvert\Bigl( \frac{1}{2}+\frac{1}{2\sqrt{2}}\Bigr)^n.
\end{equation}

For $\gamma\leq \frac{1}{2}$, we have $\bigl\lvert \mathcal{Q}_n^\gamma \bigr\rvert\leq 2^{2nh(\gamma)}$, which is shown in Sec. 1.4 of Ref. \cite{Lintbook}, where $h(\gamma)=-\gamma\log_2 \gamma - (1-\gamma)\log_2 (1-\gamma)$ is the binary entropy of $\gamma$. Thus, we have from (\ref{a3}) that
\begin{equation}
\label{a4}
p_n^\gamma\leq 2^{2nh(\gamma)}\Bigl( \frac{1}{2}+\frac{1}{2\sqrt{2}}\Bigr)^n.
\end{equation}
For $\gamma<0.015$, we have $2^{2h(\gamma)}\Bigl( \frac{1}{2}+\frac{1}{2\sqrt{2}}\Bigr)<1$; hence, $p_n^\gamma\rightarrow 0$ exponentially with $n$, which means unconditional security for $\gamma<0.015$.

\section{Unconditionally secure bit commitment from SCOT}
\label{section6}

In bit commitment, Bob commits a bit $b$ to Alice by giving her some proof of his commitment but without telling her the value of $b$, until he unveils his commitment to her. The security guarantee is twofold: Alice cannot learn $b$ before Bob unveils, and if Alice accepts Bob's unveiled bit $b$ as valid, she is guaranteed that Bob was committed to $b$ at least from some time $t'$ in the past. Similarly to the SCOT protocol of Ref. \cite{PG15.1}, the SCOT protocol presented here can be used to implement unconditionally secure relativistic bit commitment, as follows (see Fig \ref{fig1}). To commit to $b \in \{0,1\}$, Bob runs the SCOT protocol with input $b$. To unveil the committed bit $b$, Bob's agent $\mathcal{B}_b$ obtains $\bold{x}_b$ within $R_b$, and then sends $\bold{x}_b$ immediately to Alice's adjacent agent $\mathcal{A}_b$. $\mathcal{A}_b$ accepts $b$ as a valid commitment only if she receives $\bold{x}_b$ from $\mathcal{B}_b$ within the time interval $[h,h+\Delta h]$, i.e., in the boundary of $R_b$ that is adjacent to her laboratory.

It follows from the security of SCOT that this bit commitment protocol is unconditionally secure. Security against Alice follows trivially because our SCOT protocol is secure against Alice; hence, Alice does not know $b$ until Bob unveils.

Security against Bob means that if $\mathcal{A}_b$ accepts $b$ as a valid commitment then $\mathcal{A}_b$ is guaranteed that Bob was committed to $b$ from $t'$, the time coordinate of $P'$, which is the spacetime point with the greatest time coordinate (in the frame $\mathcal{F}$) that lies in the intersection of the past light cones of a point in $R_0$ and a point in $R_1$. Security against Bob follows from Minkowski causality and the monogamy of entanglement, similarly as in the security analysis of our SCOT protocol. Namely, in any cheating strategy, Bob must send in the past of $P'$ a system $B_i$ to agent $\mathcal{B}_i$, who measures the system using the information received from Alice's agent $\mathcal{A}_i$ and obtains his guess for her input $\bold{x}_i$ in $R_i$, for $i \in \{0,1\}$. The probability $p_n$, given by (\ref{4}), that Bob obtains both $\bold{x}_0$ in $R_0$ and $\bold{x}_1$ in $R_1$ is upper bounded by (\ref{5}). This means that for any cheating strategy followed by Bob, if we denote $p^i$ as the probability that Bob unveils $b=i$ successfully, for $i \in \{0,1\}$, it holds that $p^0+p^1\leq 1 + p_n$. Since $p_n\rightarrow 0$ exponentially with $n$, our bit commitment protocol is unconditionally secure against Bob.

\section{Conclusion}
\label{section7}

We presented a practical unconditionally secure SCOT protocol, where spatially separated agents need only communicate classical information, while quantum communication is only between agents at the same location, and which can be implemented with current technology. This is the first practical and unconditionally secure protocol of a cryptographic task whose unconditional security is possible in a quantum relativistic world, but impossible in a classical world, or in a quantum non-relativistic world \cite{K11.3,PG15.1}.

\begin{acknowledgments}
DPG thanks Stefano Pironio for helpful conversations. This work was begun when DPG was at the Laboratoire d'Information Quantique, Universit\'e libre de Bruxelles, with financial support from the European Union under the project QALGO, from the F.R.S.-FNRS under the project DIQIP and from the InterUniversity Attraction Poles of the Belgian Federal Government through project Photonics@be. It was completed when DPG was at IRIF, Universit\'e Paris Diderot, supported by the project ERC QCC.
\end{acknowledgments}

\appendix

\section{Proof of Eq. (\ref{15})}
\label{appa}
We show (\ref{15}) for the case $b'=0$. The proof for the case $b'=1$ follows straightforwardly by interchanging $B_0$ and $B_1$. To simplify notation, in what follows we take $\Pi_{0\bold{s}}^{\bold{r}_{0}}\equiv \Pi_{0\bold{s}0}^{\bold{r}_{0}}$, $\Pi_{1\bold{s}_{\bold{k}}}^{\bold{r}_{1}}\equiv\Pi_{1\bold{s}_{\bold{k}}0}^{\bold{r}_{1}}$, $F_{\bold{s}}\equiv F_{\bold{s}}^{0}$ and $G_{\bold{s}_{\bold{k}}}\equiv G_{\bold{s}_{\bold{k}}}^{0}$. For fixed $\bold{s}$ and $\bold{k}$, we denote by $s^j$ and $s^j_{\bold{k}}$ the $j$th entries of $\bold{s}$ and $\bold{s}_{\bold{k}}=\bold{s}\oplus\bold{k}$, respectively, and we define the sets $\tau=\bigl\{j\in[n]\big\vert s^j\neq s^j_{\bold{k}}\bigr\}$ and $\tau_c=\bigl\{j\in[n]\big\vert s^j= s^j_{\bold{k}}\bigr\}$. Let $\omega_{\bold{k}}$ be the Hamming weight of $\bold{k}$, hence, $\lvert \tau \rvert=\omega_{\bold{k}}$. Using the definitions (\ref{1}), (\ref{3}), and (\ref{18}) of the main text, we express $F_{\bold{s}}$ and $G_{\bold{s}_{\bold{k}}}$ by
\begin{eqnarray}
\label{19}
F_{\bold{s}}&=&\sum_{\bold{r}_0,\bold{r}_1 }\biggl[\bigotimes_{j\in\tau}\Bigl(\bigl( \bigl\lvert r_0^j\rangle\langle r_0^j\rvert\bigr)_{C_{s^j}^j} \otimes \bigl(\lvert \hat{r}_1^j\rangle\langle \hat{r}_1^j\rvert\bigr)_{C_{\bar{s}^j}^j} \Bigr)\times\nonumber\\
&&\quad\quad\quad\times\bigotimes_{j\in\tau_c}\Bigl(\bigl( \bigl\lvert r_0^j\rangle\langle r_0^j\rvert\bigr)_{C_{s^j}^j} \otimes \bigl(\lvert \hat{r}_1^j\rangle\langle \hat{r}_1^j\rvert\bigr)_{C_{\bar{s}^j}^j} \Bigr)\times\nonumber\\
&&\quad\quad\quad\times\bigotimes\bigl( \Pi^{\bold{r}_0}_{0\bold{s}}\bigr)_{B_0}\bigotimes\mathds{1}_{B_1}\biggr],\nonumber\\
G_{\bold{s}_{\bold{k}}}&=&\sum_{\bold{r}_0,\bold{r}_1 }\biggl[\bigotimes_{j\in\tau}\Bigl(\bigl( \bigl\lvert r_0^j\rangle\langle r_0^j\rvert\bigr)_{C_{s^j_{\bold{k}}}^j} \otimes \bigl(\lvert \hat{r}_1^j\rangle\langle \hat{r}_1^j\rvert\bigr)_{C_{\bar{s}^j_{\bold{k}}}^j} \Bigr)\times\nonumber\\
&&\quad\quad\quad\times\bigotimes_{j\in\tau_c}\Bigl(\bigl( \bigl\lvert r_0^j\rangle\langle r_0^j\rvert\bigr)_{C_{s^j_{\bold{k}}}^j} \otimes \bigl(\lvert \hat{r}_1^j\rangle\langle \hat{r}_1^j\rvert\bigr)_{C_{\bar{s}^j_{\bold{k}}}^j} \Bigr)\times\nonumber\\
&&\quad\quad\quad\times\bigotimes\mathds{1}_{B_0}\bigotimes \bigl( \Pi^{\bold{r}_1}_{1\bold{s}_{\bold{k}}}\bigr)_{B_1}\biggr].
\end{eqnarray}

Below we compute $F_{\bold{s}}G_{\bold{s}_{\bold{k}}}F_{\bold{s}}$. We express the left-hand operator $F_{\bold{s}}$ in terms of the dummy variables $\bold{r}_0$ and $\bold{r}_1$, and the right-hand one in terms of $\bold{z}_0$ and $\bold{z}_1$. The operator $G_{\bold{s}_{\bold{k}}}$ is expressed in terms of $\bold{w}_0$ and $\bold{w}_1$. From the definitions of $\tau$ and $\tau_c$, we have that $\bar{s}^j=s^j_{\bold{k}}$ and $\bar{s}^j_{\bold{k}}=s^j$ for $j\in\tau$, and $s^j_{\bold{k}}=s^j$ and $\bar{s}^j_{\bold{k}}=\bar{s}^j$ for $j\in\tau_c$. Using these properties and $\Pi^{\bold{r}_0}_{0\bold{s}}\Pi^{\bold{z}_0}_{0\bold{s}}=\delta_{\bold{r}_0,\bold{z}_0}\Pi^{\bold{r}_0}_{0\bold{s}}$, since $\bigl\{\Pi^{\bold{r}_0}_{0\bold{s}}\bigr\}_{\bold{r}_0}$ is a projective measurement, we obtain
\begin{eqnarray}
\label{20}
F_{\bold{s}}G_{\bold{s}_{\bold{k}}}F_{\bold{s}}&=&\sum_{\substack{\bold{r}_0,\bold{r}_1\\ \bold{w}_0, \bold{w}_1\\ \bold{z}_1}}
\Biggl[\bigotimes_{j\in\tau}
\biggl[\Bigl(\bigl( \bigl\lvert r_0^j\rangle\langle r_0^j\rvert\bigr)_{C_{s^j}^j} \otimes \bigl(\lvert \hat{r}_1^j\rangle\langle \hat{r}_1^j\rvert\bigr)_{C_{\bar{s}^j}^j} \Bigr)\times\nonumber\\
&&\quad\quad\times\Bigl(\bigl( \bigl\lvert \hat{w}_1^j\rangle\langle \hat{w}_1^j\rvert\bigr)_{C_{s^j}^j} \otimes \bigl(\lvert w_0^j\rangle\langle w_0^j\rvert\bigr)_{C_{\bar{s}^j}^j} \Bigr)\times\nonumber\\
&&\quad\quad\times\Bigl(\bigl( \bigl\lvert r_0^j\rangle\langle r_0^j\rvert\bigr)_{C_{s^j}^j} \otimes \bigl(\lvert \hat{z}_1^j\rangle\langle \hat{z}_1^j\rvert\bigr)_{C_{\bar{s}^j}^j} \Bigr)\biggr]\times\nonumber\\
&&\quad\quad\times\bigotimes_{j\in\tau_c}\biggl[\Bigl(\bigl( \bigl\lvert r_0^j\rangle\langle r_0^j\rvert\bigr)_{C_{s^j}^j} \otimes \bigl(\lvert \hat{r}_1^j\rangle\langle \hat{r}_1^j\rvert\bigr)_{C_{\bar{s}^j}^j} \Bigr)\times\nonumber\\
&&\quad\quad\times\Bigl(\bigl( \bigl\lvert w_0^j\rangle\langle w_0^j\rvert\bigr)_{C_{s^j}^j} \otimes \bigl(\lvert \hat{w}_1^j\rangle\langle \hat{w}_1^j\rvert\bigr)_{C_{\bar{s}^j}^j} \Bigr)\times\nonumber\\
&&\quad\quad\times\Bigl(\bigl( \bigl\lvert r_0^j\rangle\langle r_0^j\rvert\bigr)_{C_{s^j}^j} \otimes \bigl(\lvert \hat{z}_1^j\rangle\langle \hat{z}_1^j\rvert\bigr)_{C_{\bar{s}^j}^j} \Bigr)\biggr]\times\nonumber\\
&&\quad\quad\times\bigotimes\bigl( \Pi^{\bold{r}_0}_{0\bold{s}}\bigr)_{B_0}\bigotimes \bigl( \Pi^{\bold{w}_1}_{1\bold{s}_{\bold{k}}}\bigr)_{B_1}\Biggr].
\end{eqnarray}
Then we use that $\sum_{a=0}^1 \lvert a\rangle\langle a\rvert=\sum_{a=0}^1 \lvert \hat{a}\rangle\langle \hat{a}\rvert=\mathds{1}$ (the identity on a qubit) to obtain, after summing over $\bold{r}_1,\bold{w}_0,\bold{z}_1$, that
\begin{eqnarray}
\label{20.5}
F_{\bold{s}}G_{\bold{s}_{\bold{k}}}F_{\bold{s}}&=&\sum_{\bold{r}_0,\bold{w}_1}
\Biggl[\bigotimes_{j\in\tau}
\Bigl(\bigl\lvert\bigl \langle r_0^j\big\vert \hat{w}_1^j\bigr\rangle\bigr\rvert^2 \bigl( \bigl\lvert r_0^j\rangle\langle r_0^j\rvert\bigr)_{C_{s^j}^j} \otimes \mathds{1}_{C_{\bar{s}^j}^j} \Bigr)\times\nonumber\\
&&\quad\quad\times\bigotimes_{j\in\tau_c}\Bigl(\bigl( \bigl\lvert r_0^j\rangle\langle r_0^j\rvert\bigr)_{C_{s^j}^j} \otimes \bigl(\lvert \hat{w}_1^j\rangle\langle \hat{w}_1^j\rvert\bigr)_{C_{\bar{s}^j}^j} \Bigr)\times\nonumber\\
&&\quad\quad\times\bigotimes\bigl( \Pi^{\bold{r}_0}_{0\bold{s}}\bigr)_{B_0}\bigotimes \bigl( \Pi^{\bold{w}_1}_{1\bold{s}_{\bold{k}}}\bigr)_{B_1}\Biggr].
\end{eqnarray}
Since  $\bigl\lvert\bigl \langle r_0^j\big\vert \hat{w}_1^j\bigr\rangle\bigr\rvert^2=\frac{1}{2}$ and $\lvert \tau\rvert=\omega_{\bold{k}}$, it follows that
\begin{eqnarray}
\label{21}
F_{\bold{s}}G_{\bold{s}_{\bold{k}}}F_{\bold{s}}&=&\Bigl(\frac{1}{2}\Bigr)^{\omega_{\bold{k}}}\sum_{\bold{r}_0,\bold{w}_1}
\Biggl[\bigotimes_{j\in\tau}
\Bigl(\bigl( \bigl\lvert r_0^j\rangle\langle r_0^j\rvert\bigr)_{C_{s^j}^j} \otimes \mathds{1}_{C_{\bar{s}^j}^j} \Bigr)\times\nonumber\\
&&\quad\quad\times\bigotimes_{j\in\tau_c}\Bigl(\bigl( \bigl\lvert r_0^j\rangle\langle r_0^j\rvert\bigr)_{C_{s^j}^j} \otimes \bigl(\lvert \hat{w}_1^j\rangle\langle \hat{w}_1^j\rvert\bigr)_{C_{\bar{s}^j}^j} \Bigr)\times\nonumber\\
&&\quad\quad\times\bigotimes\bigl( \Pi^{\bold{r}_0}_{0\bold{s}}\bigr)_{B_0}\bigotimes \bigl( \Pi^{\bold{w}_1}_{1\bold{s}_{\bold{k}}}\bigr)_{B_1}\Biggr].
\end{eqnarray}
Using that $\bigl\{\Pi^{\bold{r}_0}_{0\bold{s}}\bigr\}_{\bold{r}_0}$ and $\bigl\{\Pi^{\bold{w}_1}_{1\bold{s}_{\bold{k}}}\bigr\}_{\bold{w}_1}$ are projective measurements, it is straightforward to see from (\ref{21}) that 
$2^{\omega_{\bold{k}}}F_{\bold{s}}G_{\bold{s}_{\bold{k}}}F_{\bold{s}}$ is a projector. Thus, $\lVert 2^{\omega_{\bold{k}}}F_{\bold{s}}G_{\bold{s}_{\bold{k}}}F_{\bold{s}}\rVert =1$, which implies (\ref{15}). The bound (\ref{5}) of the main text and security against Bob follow.

\section{Proof of bound (\ref{a3})}
\label{appb}

We show (\ref{a3}). We follow closely the proof of bound (\ref{5}). We have
\begin{eqnarray}
\label{a5}
p_n^\gamma&=&\frac{1}{2^n}\sum_{\bold{q}\in\mathcal{Q}_n^\gamma}\text{Tr}\Biggl(\sum_{\bold{s}}T_{\bold{s}} ^{\bold{q}} \Psi\Biggr)\nonumber\\
&\leq&\frac{1}{2^n} \sum_{\bold{q}\in\mathcal{Q}_n^\gamma}\Biggl\lVert \sum_{\bold{s}} T_{\bold{s}} ^{\bold{q}} \Biggr\rVert\nonumber\\
&=&\frac{1}{2^n} \sum_{\bold{q}\in\mathcal{Q}_n^\gamma}\Biggl\lVert \sum_{b'=0}^1\sum_{\bold{s}} D_{\bold{s}} ^{\bold{q}b'}\otimes R^{b'} \Biggr\rVert\nonumber\\
&=&\frac{1}{2^n} \sum_{\bold{q}\in\mathcal{Q}_n^\gamma} \max_{b'} \Biggl\lVert \sum_{\bold{s}} D_{\bold{s}} ^{\bold{q}b'} \Biggr\rVert\nonumber\\
&\leq&\frac{1}{2^n}\sum_{\bold{q}\in\mathcal{Q}_n^\gamma}\max_{b'}\sum_{\bold{k}}\max_{\bold{s}}\Bigl\lVert D_{\bold{s}} ^{\bold{q}b'}D_{\bold{s}_{\bold{k}}}^{\bold{q}b'}\Bigr\rVert,
\end{eqnarray}
where in the first line we used the linearity of the trace, in the second line we used the definition of the Schatten $\infty-$norm $\lVert \cdot \rVert$, in the third line we used (\ref{aa2}), in the fourth line we used that $\{R^0,R^1\}$ is a projective measurement, and in the last line we used Lemma \ref{lemma2} and the fact that $D_{\bold{s}} ^{\bold{q}b'}$ and $D_{\bold{s}_{\bold{k}}}^{\bold{q}b'}$ are projectors.

We define the projectors
\begin{eqnarray}
\label{a6}
F_{\bold{s}}^{\bold{q}b'}&=&\sum_{\bold{r}_0,\bold{r}_1 }\bigl( \bigl\lvert \Psi_{\bold{r}_0\bold{r}_1}^{\bold{s}}\bigr\rangle\bigl\langle\Psi_{\bold{r}_0\bold{r}_1}^{\bold{s}}\bigr\rvert\bigr)_{C}\otimes \bigl(\Pi_{0\bold{s}b'}^{\bold{r}_{b'}\oplus \bold{q}_0}\bigr)_{B_0}\otimes \mathds{1}_{B_1}\nonumber,\\
G_{\bold{s}_{\bold{k}}}^{\bold{q}b'}&=&\sum_{\bold{r}_0,\bold{r}_1}\bigl( \bigl\lvert \Psi_{\bold{r}_0\bold{r}_1}^{\bold{s}_{\bold{k}}}\bigr\rangle\bigl\langle\Psi_{\bold{r}_0\bold{r}_1}^{\bold{s}_{\bold{k}}}\bigr\rvert\bigr)_{C}\otimes  \mathds{1}_{B_0}\otimes\bigl(\Pi_{1\bold{s}_{\bold{k}}b'}^{\bold{r}_{\bar{b}'}\oplus \bold{q}_1}\bigr)_{B_1},\nonumber\\
\end{eqnarray}
which satisfy $D_{\bold{s}}^{\bold{q}b'}\leq F_{\bold{s}}^{\bold{q}b'}$ and $ D_{\bold{s}_{\bold{k}}}^{\bold{q}b'}\leq G_{\bold{s}_{\bold{k}}}^{\bold{q}b'}$. Thus, from Lemma \ref{lemma1} we have that
\begin{equation}
\label{a7}
\lVert D_{\bold{s}}^{\bold{q}b'}D_{\bold{s}_{\bold{k}}}^{\bold{q}b'}\rVert^2\leq \lVert F_{\bold{s}}^{\bold{q}b'}G_{\bold{s}_{\bold{k}}}^{\bold{q}b'}\rVert ^2=\lVert F_{\bold{s}}^{\bold{q}b'}G_{\bold{s}_{\bold{k}}}^{\bold{q}b'}F_{\bold{s}}^{\bold{q}b'}\rVert,
\end{equation}
where the equality follows from $\lVert A\rVert^2=\lVert AA^{\dagger}\rVert=\lVert A^{\dagger}A\rVert$, satisfied for any linear operator $A$ \cite{TFKW13}, and from the fact that $F_{\bold{s}}^{\bold{q}b'}$ and $G_{\bold{s}_{\bold{k}}}^{\bold{q}b'}$ are projectors.

Following the proof of (\ref{15}) in Appendix \ref{appa}, it is straightforward to show that
\begin{equation}
\label{a8}
\lVert F_{\bold{s}}^{\bold{q}b'}G_{\bold{s}_{\bold{k}}}^{\bold{q}b'}F_{\bold{s}}^{\bold{q}b'}\rVert= \Bigl(\frac{1}{2}\Bigr)^{\omega_\bold{k}},
\end{equation}
where $\omega_\bold{k}$ is the Hamming weight of $\bold{k}$. Thus, from (\ref{a5}), (\ref{a7}) and (\ref{a8}), we obtain
\begin{eqnarray}
\label{a9}
p_n^\gamma&\leq& \frac{1}{2^n}\sum_{\bold{q}\in\mathcal{Q}_n^\gamma}\sum_{\bold{k}} \Bigl(\frac{1}{\sqrt{2}}\Bigr)^{\omega_\bold{k}}\nonumber\\
&=&\frac{\lvert \mathcal{Q}_n^\gamma \rvert}{2^n}\sum_{\omega=0}^{n}\Bigl(\begin{matrix}
  n \\  \omega \end{matrix}\Bigr)\Bigl(\frac{1}{\sqrt{2}}\Bigr)^\omega\nonumber\\
&=&\lvert \mathcal{Q}_n^\gamma \rvert\Bigl(\frac{1}{2}+\frac{1}{2\sqrt{2}}\Bigr)^n,
\end{eqnarray}
which is the claimed bound (\ref{a3}).

\end{document}